\documentclass[aps,prd,twocolumn,showpacs,superscriptaddress]{revtex4-1}

\usepackage{graphicx}
\usepackage{booktabs}
\usepackage{amsmath,amssymb,bm}
\usepackage{multirow}
\usepackage{float}
\usepackage{hyperref}
\usepackage{booktabs}
\hypersetup{colorlinks=true,linkcolor=blue,citecolor=blue,urlcolor=blue}
\newcommand{\orcid}[1]{%
  \href{https://orcid.org/#1}{%
    \includegraphics[height=2ex, keepaspectratio]{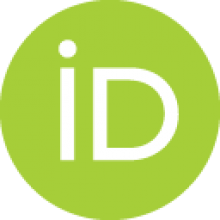}%
  }%
}

\begin{document}

\title{Robust parameter inference for Taiji via time-frequency contrastive learning and normalizing flows}

\author{Tian-Yang Sun\orcid{0009-0002-5109-6420}\email{suntianyang@stumail.neu.edu.cn}}
\affiliation{Key Laboratory of Cosmology and Astrophysics (Liaoning), College of Sciences, Northeastern University, Shenyang 110819, China}

\author{Bo Liang\orcid{0009-0002-6547-8577}\email{liangbo22@mails.ucas.ac.cn}}
\affiliation{Center for Gravitational Wave Experiment, National Microgravity Laboratory, Institute of Mechanics, Chinese Academy of Sciences, Beijing 100190, China}

\author{Ji-Yu Song\orcid{0009-0003-8111-0470}\email{songjiyu@stumail.neu.edu.cn}}
\affiliation{Key Laboratory of Cosmology and Astrophysics (Liaoning), College of Sciences, Northeastern University, Shenyang 110819, China}

\author{Song-Tao Liu\orcid{0000-0001-5966-2461}\email{songtaoliu_phy@163.com}}
\affiliation{Key Laboratory of Cosmology and Astrophysics (Liaoning), College of Sciences, Northeastern University, Shenyang 110819, China}

\author{Shang-Jie Jin\orcid{0000-0003-3697-3501}\email{jinshangjie@stumail.neu.edu.cn}}
\affiliation{Key Laboratory of Cosmology and Astrophysics (Liaoning), College of Sciences, Northeastern University, Shenyang 110819, China}

\author{He Wang\orcid{0000-0002-1353-391X}}
\thanks{Corresponding author}
\email{hewang@ucas.ac.cn}
\affiliation{Taiji Laboratory for Gravitational Wave Universe (Beijing/Hangzhou), University of Chinese Academy of Sciences (UCAS), Beijing 100049, China}
\affiliation{International Centre for Theoretical Physics Asia-Pacific (ICTP-AP), University of Chinese Academy of Sciences (UCAS), Beijing 100049, China}

\author{Ming-Hui Du\orcid{0000-0003-2155-3280}}
\thanks{Corresponding author}
\email{duminghui@imech.ac.cn}
\affiliation{Center for Gravitational Wave Experiment, National Microgravity Laboratory, Institute of Mechanics, Chinese Academy of Sciences, Beijing 100190, People’s Republic of China}

\author{Jing-Fei Zhang\orcid{0000-0002-3512-2804}\email{jfzhang@neu.edu.cn}}
\affiliation{Key Laboratory of Cosmology and Astrophysics (Liaoning), College of Sciences, Northeastern University, Shenyang 110819, China}

\author{Xin Zhang\orcid{0000-0002-6029-1933}}
\thanks{Corresponding author}
\email{zhangxin@neu.edu.cn}
\affiliation{Key Laboratory of Cosmology and Astrophysics (Liaoning), College of Sciences, Northeastern University, Shenyang 110819, China}
\affiliation{Key Laboratory of Data Analytics and Optimization for Smart Industry (Ministry of Education), Northeastern University, Shenyang 110819, China}
\affiliation{National Frontiers Science Center for Industrial Intelligence and Systems Optimization, Northeastern University, Shenyang 110819, China}

\begin{abstract}
Transient noise artifacts, commonly referred to as glitches, pose a major challenge to parameter inference for space-based gravitational-wave (GW) observations. We develop a glitch-robust amortized inference framework for massive black hole binaries in the Taiji detector configuration by combining conditional normalizing flows, a time-frequency multimodal fusion encoder, and contrastive learning. To enable large-scale training on contaminated data, we further introduce a neural glitch generator that produces high-fidelity synthetic transients at substantially reduced computational cost. Systematic experiments show that, under glitch contamination, the proposed method yields more accurate and better-calibrated posteriors than a conventional Markov Chain Monte Carlo baseline. In ablation studies, the full time-frequency model with contrastive learning performs best overall and remains robust to variations in glitch duration and merger-relative timing. We further show that standard coverage diagnostics alone are insufficient to fully assess posterior fidelity. We therefore complement them with the continuous ranked probability score, which provides a stricter assessment of global distributional agreement in non-ideal GW data. Taken together, these results establish deep-learning-based amortized inference as a promising framework for fast and robust Bayesian parameter estimation in future space-based GW observations.
\end{abstract}

\maketitle

\section{Introduction}

The direct detection of gravitational waves (GWs) has opened a new window onto the universe, enabling precision tests of general relativity in the strong-field regime~\cite{LIGOScientific:2016lio,LIGOScientific:2017adf,Sakstein:2017xjx,LIGOScientific:2018dkp,LIGOScientific:2019fpa,LIGOScientific:2020tif,LIGOScientific:2021aug,LIGOScientific:2021sio,LIGOScientific:2025jau} and providing independent probes of cosmology~\cite{Schutz:1986gp,Verde:2019ivm,DES:2019ccw,DES:2020nay,Palmese:2021mjm,Gong:2021jgg,Gong:2023ffb,Song:2025ddm,Song:2025bio}. Next-generation ground-based detectors are expected to further sharpen our understanding of the cosmic expansion history and large-scale structure~\cite{Zhao:2010sz,Berti:2015itd,Cai:2016sby,Chen:2017rfc,Wang:2018lun,Zhang:2018byx,Zhang:2019ple,Zhang:2019loq,Li:2019ajo,Zhang:2019ylr,ET:2019dnz,Jin:2020hmc,Song:2022siz,Han:2025fii,Jin:2025dvf,Du:2025odq}. Meanwhile, planned space-based missions-the Laser Interferometer Space Antenna (LISA)~\cite{LISA:2017pwj}, TianQin~\cite{TianQin:2015yph}, and Taiji~\cite{Hu:2017mde} will extend the observational frontier to the millihertz band, granting access to massive black hole binaries that are beyond the reach of ground-based facilities~\cite{Klein:2015hvg,Wang:2019tto,Zhao:2019gyk,Wang:2021srv,Jin:2021pcv,Jin:2023zhi,Jin:2023sfc,Dong:2024bvw,Dong:2025ikq,Song:2026kii,Dong:2026uxr}. Extracting the full science from these sources demands accurate Bayesian parameter estimation for source characterization and multimessenger follow-up~\cite{Marsat:2020rtl,Gao:2024uqc}, which in turn requires a thorough understanding of the detector noise environment~\cite{s11433}. A particularly pressing concern is the presence of transient instrumental artifacts, or glitches, which introduce non-Gaussian contamination into the data streams of both ground-based and space-based detectors~\cite{Powell:2015ona,Zevin:2016qwy,Edwards:2020tlp,Baghi:2021tfd,LISAPathfinder:2022awx,Spadaro:2023muy}.

Transient noise refers to short-duration, non-Gaussian features in the instrumental noise. In the context of space-based GW detectors, the current understanding of such glitches is largely informed by the LISA Pathfinder (LPF) mission. Based on LPF observations, these glitches are commonly described in terms of acceleration and displacement glitches according to their physical origin. Acceleration glitches are associated with spurious forces acting on the test mass, for example due to outgassing events, whereas displacement glitches arise from phase fluctuations in the optical measurement link~\cite{Baghi:2021tfd,LISAPathfinder:2022awx,LISAPathfinder:2022awx,Houba:2024tyn,LISAPathfinder:2024ucp}. Their occurrence rate is approximately once per day, and can reach up to three events per day under standard three-satellite operational conditions~\cite{LISAPathfinder:2022awx,Gair:2025chu}. A broader variety of transient morphologies has long been observed in ground-based detectors, which has motivated the adaptation of detector-characterization and mitigation strategies across the two communities~\cite{Powell:2015ona,Zevin:2016qwy,Zevin:2023rmt}.

The frequent presence of transient noise profoundly impacts the execution of space-based GW science missions~\cite{Robson:2018jly,Spadaro:2023muy,Gair:2025chu}. Unmodeled glitches can bias the parameter estimation of  GW signals, in severe cases leading to qualitatively incorrect posterior inferences~\cite{Spadaro:2023muy,Castelli:2024sdb,Cornish:2014kda,Cornish:2020dwh}. They can also reduce the recoverable signal-to-noise ratio and elevate the false-alarm rate across data-analysis pipelines~\cite{Baghi:2021tfd,Houba:2024tyn}. Conversely, if these transients are modeled or marginalized appropriately, they may be incorporated into more flexible descriptions of instrumental noise relevant for stochastic GW background searches~\cite{Alvey:2024uoc,Muratore:2023gxh}.

To address these non-stationary disturbances, several established approaches have been developed, many of them drawing on techniques first matured in ground-based  GW detector characterization~\cite{Cornish:2014kda,Powell:2015ona,Zevin:2016qwy}. Time-Delay Interferometry (TDI) cancellation can adapt to arbitrary unknown glitches and remains effective for unequal arm-length scenarios; however, it cannot achieve perfect cancellation and is limited by the associated loss of  GW sensitivity~\cite{Wu:2022qov,Blelly:2021oim}. The auxiliary criteria method utilizing the null T channel provides a simple mechanism to distinguish  GW signals from glitches, but it serves primarily as a discriminator rather than an active mitigation strategy~\cite{Edwards:2020tlp,Muratore:2022nbh}. Template-based or Bayesian subtraction methods can achieve accurate detection and parametrized glitch removal in waveform-controlled settings~\cite{Cornish:2014kda,Sauter:2025iey,Malz:2025xdg,Muratore:2025knh}; however, they rely on accurate waveform models and may become computationally costly or fragile when applied to poorly modeled glitch populations~\cite{Wu:2023rpn,Sauter:2025iey}. Windowing and gating methods provide a non-parametric alternative applicable to broad glitch classes, but they inevitably sacrifice portions of the signal and are less suitable for short-duration events~\cite{Pankow:2018qpo,Davis:2022ird}. In recent years, deep learning has driven a paradigm shift in complex data analysis across diverse scientific disciplines~\cite{Baldi:2014kcg,Carleo:2019ptp,Cranmer:2020wew,Karniadakis:2021,Sun:2024ywb,Sun:2025ypd} and has also shown strong performance in GW data-analysis tasks such as signal detection, noise characterization, and parameter estimation~\cite{LIGOScientific:2016gtq,LIGOScientific:2017tza,Gabbard:2017lja,George:2017fbn,George:2017qtr,Llorens-Monteagudo:2018ubm,Yan:2022spw,Wang:2023lif,Wang:2024oei,Ashton:2025jhn}. In the context of noise mitigation, deep-learning-based denoising or subtraction methods offer substantial speed advantages and strong adaptability, but many of these approaches inherently lack calibrated posterior uncertainty quantification~\cite{Mohanty:2023mjn,Houba:2024tyn,Xu:2024jbo,Mogushi:2021cpw,Dax:2021tsq}.

Motivated by these limitations, we develop a robust end-to-end parameter-inference framework based on neural posterior estimation (NPE). NPE and related simulation-based inference techniques have recently been widely adopted in  GW data analysis due to their ability to rapidly generate accurate posteriors, bypassing the computational bottlenecks of traditional sampling algorithms~\cite{Dax:2021tsq,Dax:2022pxd,Langendorff:2022fzq,Shih:2023jme,Leyde:2023iof,Dax:2024mcn,Santoliquido:2025lot,Chan:2025kyu,Santoliquido:2025aiq,Qin:2025mvj,Caldarola:2025oxr,Du:2025xdq,Kofler:2025dux,Liu:2026nhu,Spadaro:2026evb,Zhang:2026okw}. Building upon these foundations, our approach utilizes an amortized inference framework parameterized by normalizing flows~\cite{Sun:2023vlq}. Crucially, to ensure robustness against non-ideal transient contamination, we introduce a time-frequency multimodal fusion encoder integrated with a contrastive learning objective~\cite{Chen:2020simclr,Radford:2021clip,Liu:2023ssl,Wang:2026ccsn}. By leveraging this contrastive paradigm, the proposed architecture learns latent representations that are less sensitive to nuisance contamination, thereby enabling more robust conditional posterior estimation. This framework alleviates the computational burden of traditional sampling-based pipelines and enables rapid posterior estimation. By training directly on glitch-contaminated simulations, it is designed to improve robustness to unmodeled transient artifacts and to reduce the bias induced by likelihood misspecification.

\section{Methodology}

\begin{figure*}[htbp]
    \centering
    \includegraphics[width=0.9\textwidth]{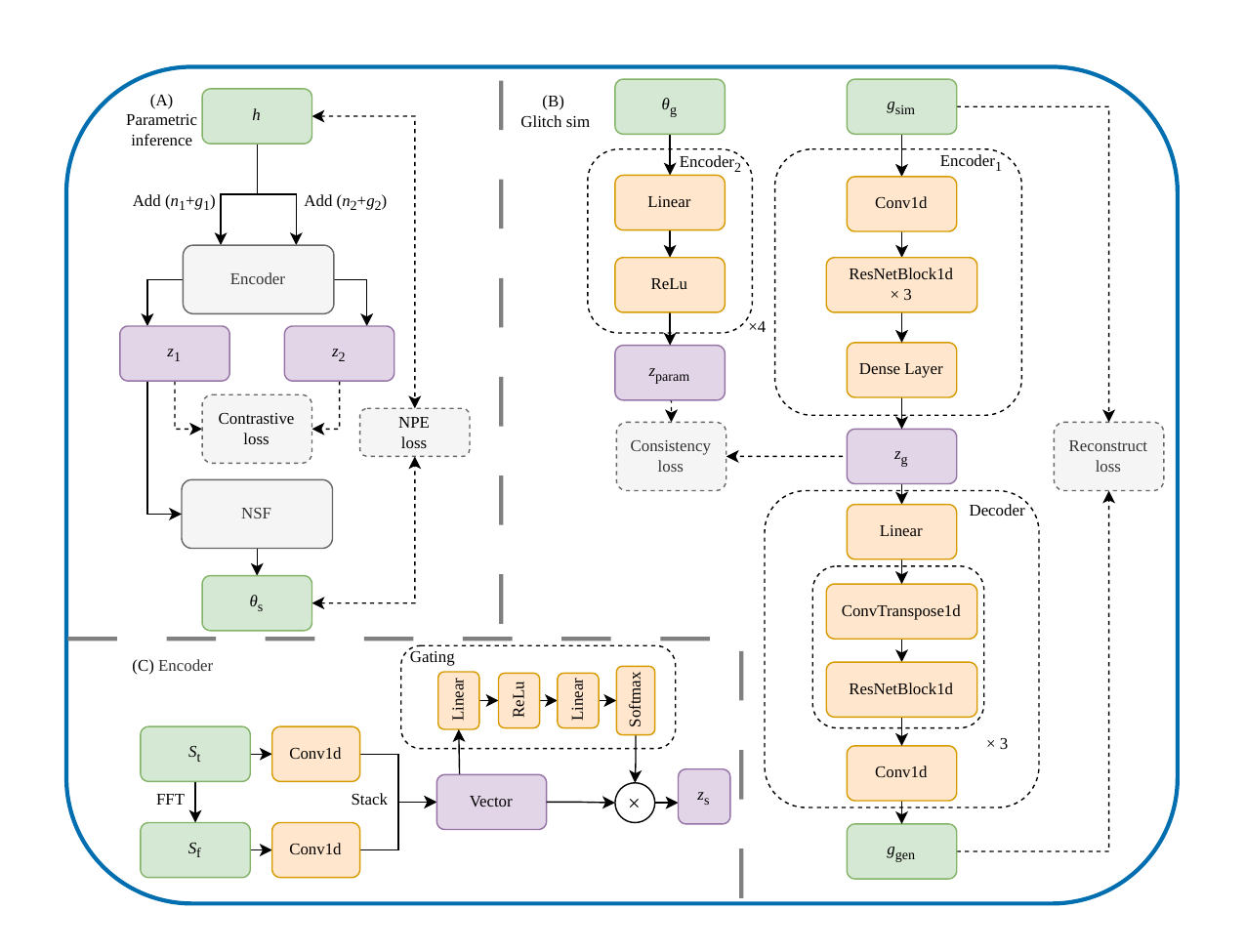}
    \caption{Overall architecture of the proposed glitch-robust inference framework. Panel (A) shows the posterior inference module, in which two independently contaminated realizations of the same source are encoded to construct a contrastive objective and to provide conditioning context for a normalizing flow. Panel (B) shows the neural glitch generator, which maps phenomenological glitch descriptors and latent variables to synthetic transient realizations and is trained with reconstruction and consistency losses. Panel (C) shows the time-frequency multimodal fusion encoder: the input time series and its Fourier-domain representation are processed by separate convolutional branches and combined by a learned gating module.}
    \label{fig:architecture}
\end{figure*}

\subsection{Data simulation and training dataset}
\label{subsec:data_simulation}

We simulate the observational environment according to the projected design specifications of a Taiji detector configuration. The Taiji constellation consists of three drag-free spacecraft arranged in an equilateral triangle with an arm length of approximately $3\times10^{6}\,\mathrm{km}$, yielding high sensitivity in the frequency band $[10^{-4},\,1]~\mathrm{Hz}$. The target sources considered in this work are massive black hole binary (MBHB) coalescences, modeled using the 11-parameter inspiral-merger-ringdown phenomenological waveform approximant IMRPhenomT, which describes the inspiral, merger, and ringdown stages. The source parameter vector is written as
\begin{equation}
\theta = \left(\mathcal{M}_{c,z}, q, \chi_{z,1}, \chi_{z,2}, t_c, \varphi_c, D_L, \iota, \lambda, \beta, \psi\right),
\end{equation}
where $\mathcal{M}_{c,z}$ is the redshifted chirp mass, $q$ is the mass ratio, $\chi_{z,1}$ and $\chi_{z,2}$ are the aligned spin components, $t_c$ and $\varphi_c$ denote the coalescence time and phase, $D_L$ is the luminosity distance, $\iota$ is the inclination angle, $\lambda$ and $\beta$ are the ecliptic longitude and latitude, and $\psi$ is the polarization angle.

The prior distributions are chosen to be broad and astrophysically motivated, so as to cover the relevant MBHB parameter space while remaining consistent with the Taiji simulation setup adopted in this work. Specifically, $\mathcal{M}_{c,z}$ and $D_L$ are sampled log-uniformly, the angular variables are sampled isotropically where appropriate, and the remaining intrinsic parameters are drawn from uniform distributions within prescribed ranges. The complete prior settings are summarized in Table~\ref{tab:data_priors}.

The simulated detector data are modeled as
\begin{equation}
s = h(\theta) + n + g,
\label{eq:data_model}
\end{equation}
where $h(\theta)$ denotes the deterministic GW response of the Taiji detector configuration, including the corresponding TDI response, $n$ represents stationary Gaussian instrumental noise characterized by the detector power spectral density, and $g$ denotes a transient non-stationary glitch component. Relative to the standard glitch-free setting, the inclusion of $g$ introduces localized excess power and morphological distortions that cannot be described by stationary Gaussian noise alone. The resulting simulations therefore provide a more realistic training and evaluation setting for practical analysis scenarios. For the glitch component $g$, we adopt an LPF-inspired legacy acceleration-glitch model. In this model, the glitch is parameterized by an impulse scale $dv$ and two characteristic timescales, $\tau_{\rm 1}$ and $\tau_{\rm 2}$, and is introduced at the acceleration level before being integrated into the corresponding interferometric observable. We sample these parameters over the ranges $dv \in [1\times10^{-15},\,1\times10^{-13}]$, $\tau_{\rm 1} \in [2.0,\,60.0]$, and $\tau_{\rm 2} \in [2.0,\,60.0]$, which define the family of short-duration transient artifacts considered in this work~\cite{Armano:2018kix,Baghi:2021tfd}. During training, the glitch amplitude is further rescaled such that the glitch signal-to-noise ratio spans the range $[8,\,32]$, thereby exposing the inference network to a broad set of moderate contamination levels. For evaluation, we additionally consider substantially stronger glitch realizations, with test cases extending to glitch SNRs as high as $10^3$, in order to assess the extrapolative robustness of the proposed framework under extreme non-stationary contamination.

To preserve complementary information from both the astrophysical signal and the transient contamination, each simulated sample is represented in both the time and frequency domains. This dual-domain construction allows the training set to capture waveform morphology, spectral structure, and glitch-induced non-stationarity simultaneously, while preserving flexibility for the downstream inference model. Clean-signal and glitch-contaminated realizations are generated over the same astrophysical parameter space, ensuring that the effect of transient contamination is learned against a consistent physical background.

Based on this simulation strategy, the training data are generated online during each training epoch. In each epoch, $10^5$ statistically independent samples are simulated over the same astrophysical parameter space, of which $90\%$ are used for training and the remaining $10\%$ for validation. The source parameters are sampled using Latin hypercube sampling to improve the coverage efficiency of the high-dimensional parameter space. The test set is generated separately after training is completed and remains strictly non-overlapping with the training and validation data. The training subset is used to optimize the inference model, the validation subset is used for hyperparameter tuning and convergence monitoring, and the test subset is reserved exclusively for performance evaluation.

\begin{table*}[htbp]
\centering
\caption{Prior ranges adopted for the simulated MBHB source parameters in the Taiji detector configuration.}
\label{tab:data_priors}
\begin{tabular}{cccc}
\toprule
\hline
Parameter & Symbol & Prior range & Sampling rule \\
\midrule
\hline
Redshifted chirp mass & $\mathcal{M}_{c,z}$ & $[10^{5.5},\,10^{6.5}]\,M_\odot$ & log-uniform \\
Mass ratio & $q$ & $[0.1,\,0.9]$ & uniform \\
Primary aligned spin & $\chi_{z,1}$ & $[-0.9,\,0.9]$ & uniform \\
Secondary aligned spin & $\chi_{z,2}$ & $[-0.9,\,0.9]$ & uniform \\
Coalescence time & $t_c$ & $[t_{\mathrm{GPS}}-0.01,\,t_{\mathrm{GPS}}+0.01]~\mathrm{day}$ & uniform \\
Coalescence phase & $\varphi_c$ & $[0,\,2\pi]$ & uniform \\
Luminosity distance & $D_L$ & $[10^{4},\,10^{5}]~\mathrm{Mpc}$ & log-uniform \\
Inclination angle & $\iota$ & $[0,\,\pi]$ & uniform in $\cos\iota$ \\
Ecliptic longitude & $\lambda$ & $[0,\,2\pi]$ & uniform \\
Ecliptic latitude & $\beta$ & $[-\pi/2,\,\pi/2]$ & uniform in $\sin\beta$ \\
Polarization angle & $\psi$ & $[0,\,\pi]$ & uniform \\
\bottomrule
\hline
\end{tabular}
\end{table*}

\subsection{Framework}
\label{subsec:framework}

Figure~\ref{fig:architecture} presents the overall framework proposed in this work. Our method is designed for robust parameter inference of MBHB signals in the presence of non-stationary instrumental glitches. The full system consists of two complementary components: a glitch-robust posterior inference module and a neural glitch generator. The former learns an amortized approximation to the conditional posterior distribution of the source parameters from glitch-contaminated detector data, while the latter serves as a fast surrogate for synthesizing transient artifacts during training.

To this end, the inference module combines a conditional normalizing flow with a time-frequency multimodal encoder, so that both temporal and spectral representations of the glitch-contaminated detector data can be exploited within a unified inference pipeline. The time- and frequency-domain features are fused through a learned gating operator that adaptively reweights the two representations for each contaminated realization. This design follows the general idea of attention-based multi-view fusion in  GW data analysis and multimodal deep learning~\cite{10.1109/ICASSP.2017.7952693,Zhao:2023tqr,Wu:2024tpr,Al-Shammari:2026rvd}.

A key challenge in this setting is the efficient generation of diverse glitch-contaminated training samples. Direct physics-based simulation of realistic instrumental transients is computationally expensive and unsuitable for sustained large-scale online training. To overcome this bottleneck and enable the online generation of diverse contaminated samples, we therefore introduce a neural glitch generator that is trained offline to reproduce the morphology and parametric variability of physically simulated glitches. Once trained, this surrogate is embedded into the data-generation pipeline and provides fast transient synthesis throughout posterior-network training.

Under this framework, the glitch-robust posterior inference module and the neural glitch generator play distinct but complementary roles. The inference module is responsible for extracting source-relevant information and estimating posterior densities from glitch-contaminated detector data, whereas the neural glitch generator provides a scalable source of nuisance realizations that exposes the inference network to a broad variety of non-stationary artifacts. The internal structure of the time-frequency multimodal fusion encoder is shown in Panel~(C) of Fig.~\ref{fig:architecture}, where time-domain and frequency-domain features are fused by a learned gating operator, while Panels~(A) and~(B) summarize the inference and generation branches, respectively.

\subsection{Glitch-robust posterior inference module}
\label{subsec:posterior_module}

To construct a flexible and tractable approximation to the posterior distribution $q_{\phi}(\theta\mid x)$, the inference module adopts a conditional normalizing flow, as illustrated in Panel~(A) of Fig.~\ref{fig:architecture}. Normalizing flows define a sequence of invertible and differentiable transformations that map a simple base distribution $p_0(z)=\mathcal{N}(0,I)$ in $\mathbb{R}^{d}$ to a complex target density. Conditioning is introduced through a context vector $c = E_{\psi}(x)$ extracted by the encoder network, such that
\begin{equation}
    \theta = f_{\phi}(z;\,c), \qquad z \sim p_0(z).
\end{equation}
By the change-of-variables theorem, the corresponding conditional density is
\begin{equation}
    q_{\phi}(\theta\mid x)
    = p_0(z)\left|\det\frac{\partial f^{-1}_{\phi}(\theta;\,c)}{\partial \theta}\right|.
\end{equation}
In practice, the flow layers are implemented using neural spline transformations with monotonic rational-quadratic splines, and permutation layers are inserted between coupling blocks to promote full interaction among latent dimensions while retaining efficient density evaluation~\cite{DBLP:conf/nips/DurkanB0P19}.

The quality of the posterior approximation depends critically on the feature representation learned from the contaminated data.  GW signals and instrumental glitches exhibit complementary signatures in different domains: the time domain emphasizes transient onsets, burst-like morphology, and amplitude envelopes, whereas the frequency domain captures chirping spectral tracks and long-duration waveform evolution. To exploit both types of information, we adopt a time-frequency multimodal fusion encoder, shown in Panel~(C) of Fig.~\ref{fig:architecture}. The time-domain branch extracts a feature vector $c_t = E_t(x_t)$ from the contaminated time series, while the frequency-domain branch extracts a spectral feature vector $c_f = E_f(x_f)$ from the discrete Fourier transform of the same input, represented through real and imaginary channels.

The two feature branches are integrated by a gated fusion operator $\Gamma$~\cite{DBLP:conf/iclr/OvalleSMG17,Zhang:2022fwq},
\begin{equation}
    c = \Gamma(c_t,\,c_f),
    \label{eq:fusion_revised}
\end{equation}
which adaptively balances temporal and spectral information according to the morphology of the input realization. Specifically, the fusion gate is written as
\begin{equation}
    \alpha = \sigma\!\bigl(W_g[c_t;\,c_f] + b_g\bigr),
    \qquad
    c = \alpha \odot c_t + (1-\alpha)\odot c_f,
    \label{eq:gating_revised}
\end{equation}
where $W_g$ and $b_g$ are trainable parameters, $\sigma(\cdot)$ denotes the sigmoid activation, and $\odot$ is element-wise multiplication. This mechanism allows the network to dynamically emphasize the more informative modality for a given glitch-contaminated sample without requiring manual selection rules.

To further improve robustness against non-stationary contamination, we introduce an auxiliary contrastive learning objective on the latent context vectors. For a fixed source parameter vector $\theta$, two contaminated realizations sharing the same deterministic waveform response $h(\theta)$ but differing in their noise and glitch realizations form a positive pair for contrastive learning. The encoder is encouraged to map such pairs to nearby latent representations, while separating representations associated with different sources. This is implemented through an InfoNCE-style contrastive loss,
\begin{equation}
    \mathcal{L}_{\mathrm{CL}}
    = -\log
    \frac{\exp\!\bigl(\mathrm{sim}(c^{(1)},c^{(2)})/\tau\bigr)}
         {\displaystyle\sum_{j=1}^{K}\exp\!\bigl(\mathrm{sim}(c^{(1)},c^{(j)})/\tau\bigr)},
    \label{eq:cl_loss_revised}
\end{equation}
where $\mathrm{sim}(\cdot,\cdot)$ denotes cosine similarity, $\tau$ is a learnable temperature parameter, and the denominator sums over the mini-batch. By maximizing similarity between positive pairs while suppressing similarity to negatives, the encoder is driven to retain source-relevant information and discard nuisance features induced by transient contamination.

The glitch-robust posterior inference module is trained end-to-end by minimizing the negative conditional log-likelihood,
\begin{equation}
    \mathcal{L}_{\mathrm{NPE}}
    = -\mathbb{E}_{(x,\theta)}\!\bigl[\log q_{\phi}(\theta\mid x)\bigr],
    \label{eq:npe_loss_revised}
\end{equation}
which is equivalent to minimizing the forward Kullback-Leibler divergence between the true posterior and its amortized approximation. The full training objective for the inference module is therefore
\begin{equation}
    \mathcal{L}_{\mathrm{post}}
    = \mathcal{L}_{\mathrm{NPE}}
    + \lambda_{\mathrm{CL}}\,\mathcal{L}_{\mathrm{CL}},
    \label{eq:posterior_loss_revised}
\end{equation}
where $\lambda_{\mathrm{CL}}$ controls the strength of the contrastive regularization and is selected using the validation set. This objective ensures that the model learns both accurate posterior densities and glitch-invariant latent representations. 

\subsection{Neural glitch generator}
\label{subsec:glitch_module}

A major bottleneck in training glitch-robust posterior estimators is the large-scale generation of realistic glitch-contaminated data segments in the time domain. Conventional physics-based simulation of space-based detector transients typically relies on highly parameterized phenomenological models or computationally intensive transfer-function calculations, making the online production of millions of distinct glitch realizations prohibitively expensive. To overcome this limitation, we train a neural glitch generator that acts as a high-fidelity surrogate model, as illustrated in Panel~(B) of Fig.~\ref{fig:architecture}.

The generator defines a differentiable mapping from a low-dimensional continuous latent code $z_g \in \mathbb{R}^{d_g}$ to a synthetic transient strain realization $\hat{g}$. In place of expensive physical modeling at runtime, the surrogate produces glitches through efficient forward passes through convolutional and dense layers. The model is trained offline on a curated set of physically simulated instrumental transients that captures the characteristic temporal morphologies and spectral energy distributions of realistic detector anomalies. Following the design philosophy of Refs.~\cite{Sun:2025afb,Sun:2025ypd}, we adopt a simplified high-throughput architecture that prioritizes generation efficiency while preserving sufficient physical fidelity for posterior-network training.

In addition to the latent code $z_g$, the generator is conditioned on a small set of phenomenological glitch descriptors, such as $dv$, $\tau_1$ and $\tau_2$. This conditioning improves controllability over the generated transients and helps preserve correspondence with physically interpretable glitch properties. The generator is optimized using a composite objective,
\begin{equation}
    \mathcal{L}_{\mathrm{glitch}}
    = \lambda_{\mathrm{rec}}\,\mathcal{L}_{\mathrm{rec}}
    + \lambda_{\mathrm{cons}}\,\mathcal{L}_{\mathrm{cons}},
    \label{eq:glitch_loss_revised}
\end{equation}
where $\mathcal{L}_{\mathrm{rec}}$ is the mean absolute error between the generated transient $\hat{g}$ and the corresponding physically simulated reference glitch $g$, and $\mathcal{L}_{\mathrm{cons}}$ is a consistency term that penalizes deviations between the latent representation and the prescribed phenomenological glitch parameters. The coefficients $\lambda_{\mathrm{rec}}$ and $\lambda_{\mathrm{cons}}$ are tuned to balance waveform fidelity against parametric coherence during training.

Once trained, the neural glitch generator produces new transient realizations in a single forward pass and can therefore be embedded efficiently into the online simulation pipeline used for posterior-network training. Its role is not to perform parameter inference directly, but to provide a scalable and physically informed source of nuisance realizations that exposes the posterior model to a broad and diverse set of glitch morphologies. 
 
\subsection{Evaluation metrics}
\label{subsec:metrics}

The performance of the inference framework is evaluated using a set of complementary statistical metrics that probe point-estimation accuracy, posterior uncertainty, calibration, and overall distributional fidelity.
 
For each source parameter $\theta_k$ ($k=1,\ldots,11$), we report the posterior mean offset $\Delta\mu_k = \hat{\mu}_k - \theta_k^{\star}$, where $\hat{\mu}_k$ is the mean of the marginalized posterior and $\theta_k^{\star}$ is the true injected value, together with the posterior width $\sigma_k$ quantifying credible-interval breadth.

The P-P plot provides a global calibration diagnostic by measuring the empirical coverage of the recovered posteriors across an ensemble of simulated injections.
For each injection $i$ and parameter $k$, we compute the percentile rank $\pi_{ik}$ of the true value $\theta_k^{\star}$ within the one-dimensional marginalized posterior.
Under a perfectly calibrated posterior, $\{\pi_{ik}\}$ follows a uniform distribution on $[0,1]$.
We evaluate statistical consistency between the empirical cumulative distribution function (CDF) of $\{\pi_{ik}\}$ and the theoretical uniform CDF using the Kolmogorov-Smirnov (KS) test.
The KS statistic is defined as
\begin{equation}
    D_{\mathrm{KS}}
    = \sup_{t\in[0,1]}\bigl|\hat{F}(t) - t\bigr|,
\end{equation}
where $\hat{F}(t)$ is the empirical CDF of the percentile ranks.
A high p-value ($p > 0.05$) indicates that the posteriors are statistically well-calibrated, whereas a small p-value reveals systematic bias or miscalibration.
This test is applied separately to each parameter, and the resulting calibration statistics are summarized across the full set of 11 source parameters.

Standard marginal P-P coverage tests suffer from inherent degeneracies when evaluating biased posterior structure (see Appendix~\ref{app:coverage_limitations}).
To assess holistic distributional agreement, we employ the Continuous Ranked Probability Score (CRPS)~\cite{DBLP:conf/copa/VyuginT19,arnold2024decompositions}, a strictly proper scoring rule that evaluates the integrated squared difference between the posterior CDF $F(z)$ and the Heaviside indicator centered at the true parameter value $y$:
\begin{equation}
    \mathrm{CRPS}(F,\,y)
    = \int_{-\infty}^{\infty}\!\bigl(F(z) - \mathbb{I}\{z\ge y\}\bigr)^{2}\,\mathrm{d}z.
    \label{eq:crps}
\end{equation}
The CRPS simultaneously penalizes systematic mean biases and miscalibrated uncertainty widths, providing a unified measure of distributional accuracy that is sensitive to both first- and second-order moment errors~\cite{arnold2024decompositions,DBLP:journals/tmlr/FerrerR25}.
An aggregated, mean normalized CRPS score is computed over the full 11-parameter space to enable high-level architectural comparisons and systemic robustness investigations.

\section{Results and Discussion}
\label{sec:results}

\subsection{Performance of the neural glitch generator}
\label{subsec:results_generator}

\begin{figure}[t]
    \centering
    \includegraphics[width=0.48\textwidth]{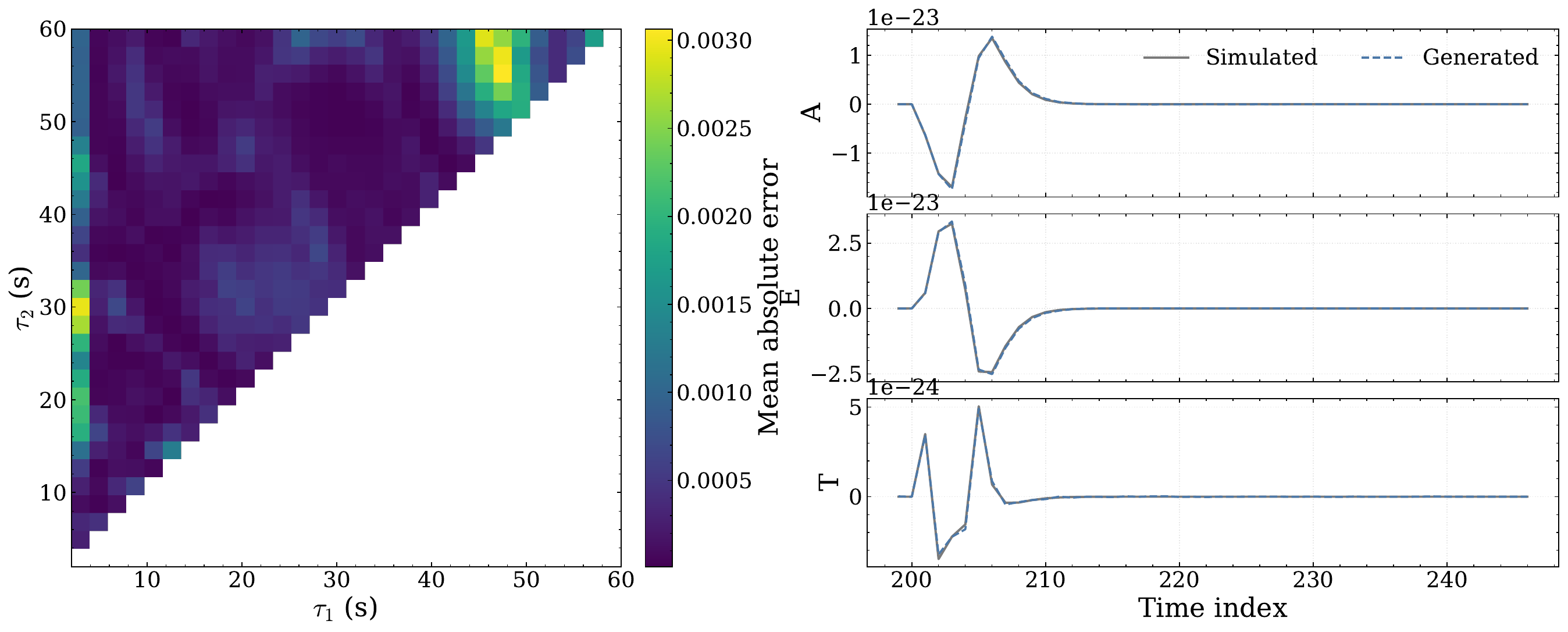}
    \caption{Fidelity of the neural glitch generator across the explored transient parameter space. The left panel shows the mean absolute reconstruction error as a function of the glitch temporal parameters $\tau_1$ and $\tau_2$, where the color scale represents the reconstruction error magnitude. The right panels compare representative physically simulated glitches and the corresponding neural-generated transients in the A, E, and T TDI channels as functions of time index.}
    \label{fig:glitchgen}
\end{figure}

A key prerequisite for training a glitch-robust posterior estimator is the ability to generate large numbers of realistic contaminated samples efficiently. We therefore evaluate the neural glitch generator separately before analyzing its downstream impact on posterior inference. Figure~\ref{fig:glitchgen} compares physically simulated glitches with the corresponding outputs of the neural surrogate across the explored temporal parameter range. The generated transients reproduce the overall morphology of the target glitches with high fidelity, including the main amplitude evolution and waveform structure relevant for contamination modeling.

The agreement in Fig.~\ref{fig:glitchgen} indicates that the surrogate captures the dominant features of the simulated glitch family rather than merely approximating their average profile. This is important because posterior robustness depends on exposing the inference network to a sufficiently rich set of contaminating morphologies during training. If the generator were only able to reproduce a narrow subset of transient structures, the downstream model would risk overfitting to a simplified nuisance distribution and failing under more diverse contamination patterns. The close correspondence between the neural-generated and physically simulated glitches therefore provides evidence that the surrogate is adequate for large-scale training use.

The quantitative reconstruction behavior shown in Fig.~\ref{fig:glitchgen} further supports this conclusion. In addition to waveform fidelity, the neural surrogate also provides a substantial computational advantage over direct physical glitch simulation. The computational cost benchmark, measured on an Intel(R) Xeon(R) w9-3495X CPU using 100 CPU cores, is summarized in Table~\ref{tab:glitch_speed}. The neural generator reduces the cost of producing large contaminated datasets by a large factor, which is essential for on-the-fly training over $10^5$ samples per epoch. The reconstruction error remains low throughout the scanned range of temporal parameters, indicating that the generator maintains stable fidelity across a nontrivial region of the glitch parameter space rather than only near isolated examples. This stability is precisely what is required for on-the-fly data generation during posterior-network training, where millions of contaminated realizations must be synthesized efficiently without repeatedly invoking expensive physical simulations.

\begin{table}[t]
\centering
\caption{Computational cost comparison between direct physical glitch simulation and the neural glitch generator.}
\label{tab:glitch_speed}
\begin{tabular}{lccc}
\toprule
\hline
Method & Time for $10^5$ samples & Speedup \\
\hline
\midrule
Physical simulation& 67.56~h & $1\times$ \\
Neural generator & 2~min & $795\times$ \\
\hline
\bottomrule
\end{tabular}
\end{table}

From a methodological perspective, the role of the neural glitch generator is not to replace physical modeling as the ultimate description of instrumental artifacts, but to replace it as a computational bottleneck. By substituting numerically expensive transient simulation with a lightweight neural forward model, the generator makes it feasible to train the inference module on a large and morphologically diverse contaminated dataset. Figure~\ref{fig:glitchgen} therefore establishes that the generator is not merely an auxiliary module, but plays an important practical role in scaling the overall pipeline.

\subsection{Robust posterior recovery under glitch contamination}
\label{subsec:results_inference}

\begin{figure*}[t]
    \centering
    \includegraphics[width=0.88\textwidth]{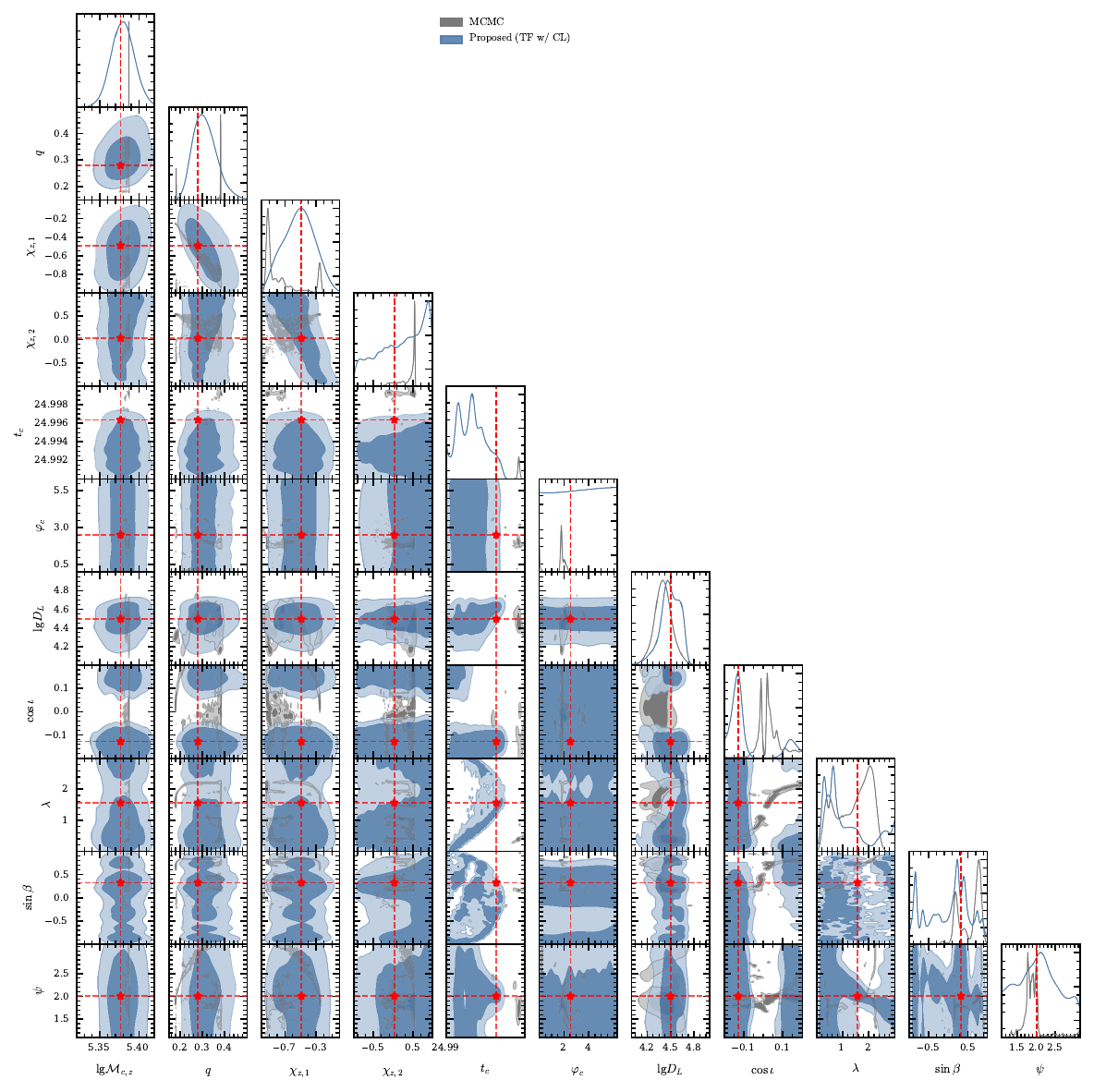}
    \caption{Representative multidimensional posterior distributions for a glitch-contaminated injection. The corner plot compares the conventional Bayesian MCMC baseline with the proposed time-frequency normalizing-flow model with contrastive learning. Diagonal panels show the one-dimensional marginalized posteriors for selected source parameters, and off-diagonal panels show the corresponding pairwise joint posterior contours. Vertical and horizontal reference markers denote the injected parameter values. The dark area and the light area respectively represent the $1~\sigma$ confidence interval and the $2~\sigma$ confidence interval.}
    \label{fig:triangle}
\end{figure*}

We first assess whether the proposed framework can recover accurate and well-calibrated posteriors from glitch-contaminated data. Figure~\ref{fig:triangle} presents representative multidimensional posterior distributions obtained under non-stationary contamination. For the proposed time-frequency flow equipped with contrastive learning, the recovered posterior contours remain closely aligned with the injected source parameters and preserve a physically reasonable correlation structure. By contrast, under the same contaminated conditions, the conventional Markov Chain Monte Carlo (MCMC) baseline exhibits visible displacements of posterior support away from the injected values for several parameters, indicating that likelihood misspecification induced by the transient artifact leads to biased inference.

This difference can be seen both in the multidimensional posterior contours in Figure~\ref{fig:triangle} and in the marginalized posterior summaries reported in Table~\ref{tab:posterior_summary}, and can be understood from the way the two approaches handle transient contamination. The conventional MCMC framework evaluates a Whittle likelihood derived under the assumption of stationary Gaussian noise. When an unmodeled glitch is present, this assumption is no longer strictly valid, and the resulting likelihood misspecification can distort posterior contours and shift credible regions away from the injected values. By contrast, the proposed amortized inference framework is trained directly on glitch-contaminated simulations over a broad range of nuisance realizations. As a result, the encoder and flow learn source-relevant features that remain stable under transient contamination, rather than relying on a strictly stationary likelihood model. The alignment observed in Fig.~\ref{fig:triangle} therefore suggests that the proposed model has learned a representation of glitch-contaminated data that generalizes well beyond the nominal training regime and remains substantially more robust to non-stationary noise than the conventional MCMC baseline. 

\begin{table}[htbp]
\centering
\caption{Marginalized posterior summaries for the glitch-contaminated injection shown in Figure~\ref{fig:triangle}. For each parameter, we report the posterior median together with the corresponding asymmetric $1\,\sigma$ credible interval for the conventional MCMC baseline and the proposed NPE model.}
\label{tab:posterior_summary}
\begin{tabular}{lccc}
\toprule
\hline
Parameter & True Value & MCMC baseline & NPE \\
\midrule
\hline
$\log \mathcal{M}_{c,z}$ 
& $5.38$
& $5.38_{-4.22\times10^{-4}}^{+3.19\times10^{-5}}$
& $5.37_{-1.47\times10^{-2}}^{+1.49\times10^{-2}}$ \\

$q$
& $0.28$
& $0.37_{-0.19}^{+2.08\times10^{-3}}$
& $0.30_{-4.84\times10^{-2}}^{+5.90\times10^{-2}}$ \\

$\chi_{1,z}$
& $-0.49$
& $-0.87_{-5.85\times10^{-2}}^{+0.60}$
& $-0.52_{-0.23}^{+0.20}$ \\

$\chi_{2,z}$
& $0.03$
& $0.48_{-0.38}^{+4.82\times10^{-2}}$
& $0.28_{-0.76}^{+0.53}$ \\

$t_c$
& $24.99$
& $25.00_{-2.48\times10^{-3}}^{+4.50\times10^{-4}}$
& $24.99_{-2.61\times10^{-3}}^{+2.07\times10^{-3}}$ \\

$\varphi_c$
& $2.52$
& $1.79_{-1.79}^{+0.25}$
& $3.20_{-2.18}^{+2.13}$ \\

$\log D_L$
& $4.50$
& $4.40_{-8.95\times10^{-2}}^{+9.54\times10^{-2}}$
& $4.49_{-0.12}^{+0.13}$ \\

$\cos\iota$
& $-0.13$
& $0.02_{-0.39}^{+0.18}$
& $-0.14_{-0.36}^{+0.45}$ \\

$\lambda$
& $1.57$
& $1.57_{-1.16}^{+0.62}$
& $1.42_{-1.01}^{+3.92}$ \\

$\sin\beta$
& $0.36$
& $0.66_{-0.50}^{+0.14}$
& $0.12_{-0.82}^{+0.39}$ \\

$\psi$
& $2.00$
& $1.87_{-0.15}^{+0.14}$
& $1.50_{-1.07}^{+0.95}$ \\

\bottomrule
\hline
\end{tabular}
\end{table}

\begin{figure}[t]
    \centering
    \includegraphics[width=0.48\textwidth]{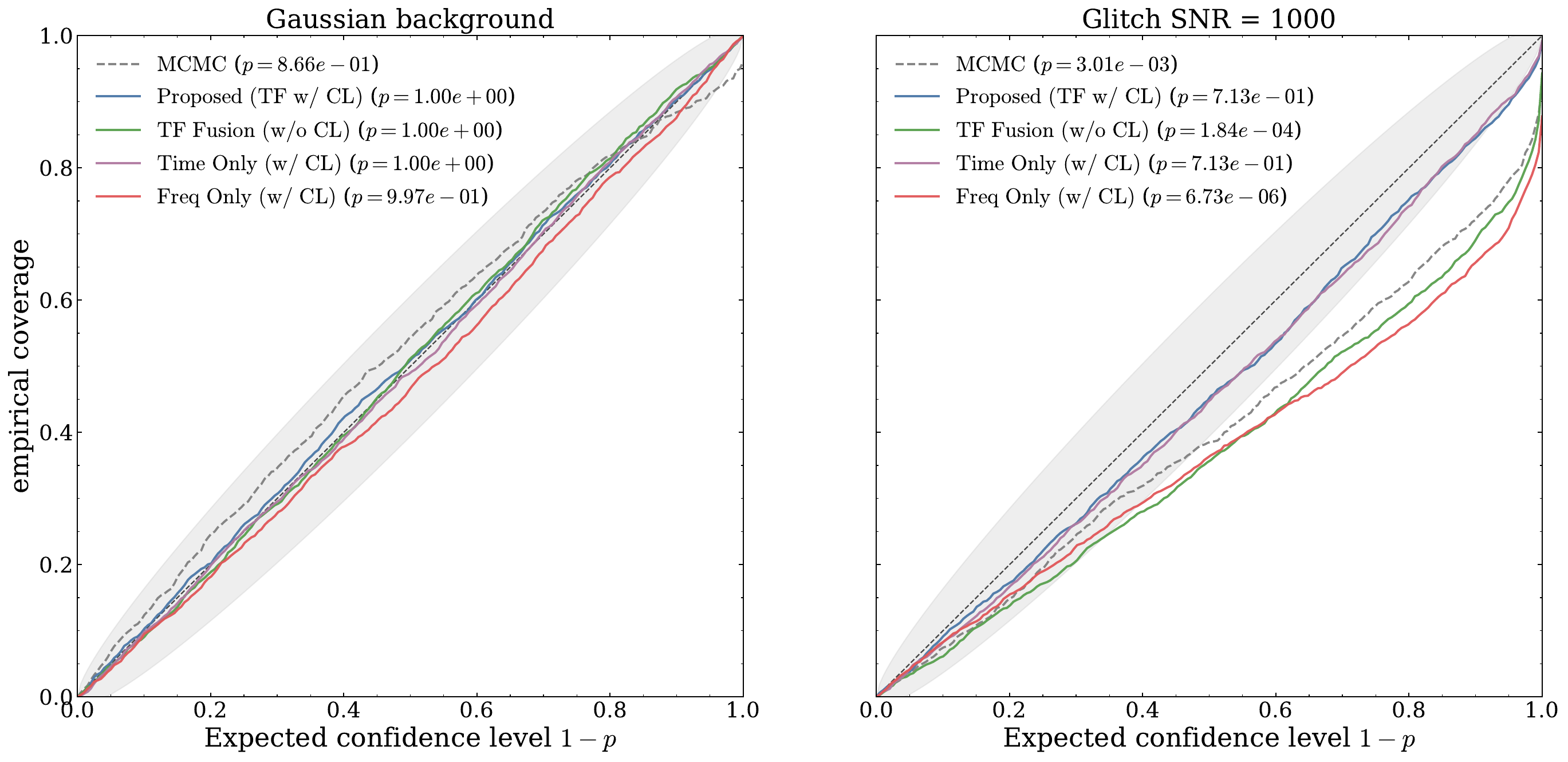}
    \caption{P-P plots for posterior calibration in glitch-free and glitch-contaminated regimes. The horizontal axis shows the nominal credibility level, and the vertical axis shows the empirical coverage fraction. Different curves correspond to the MCMC baseline and the neural inference variants, including the full time-frequency model with contrastive learning and its reduced ablations.}
    \label{fig:ppplot}
\end{figure}

We next examine whether this qualitative improvement is accompanied by reliable uncertainty quantification. The P-P plots in Fig.~\ref{fig:ppplot} provide a direct calibration test by comparing empirical coverage against the nominal confidence level. In the glitch-free regime, both the MCMC baseline and all neural inference models exhibit satisfactory calibration, consistent with the fact that these methods operate close to their ideal conditions when the data contain only stationary Gaussian noise. Under elevated glitch contamination, however, the difference becomes pronounced. The MCMC baseline departs visibly from the diagonal and fails the calibration test, whereas the proposed time-frequency model with contrastive learning remains close to the expected coverage curve and retains a high p-value.

Figure~\ref{fig:ppplot} also provides an early ablation view of the model calibration behavior. In the contaminated regime, removing contrastive learning from the time-frequency model leads to a strong degradation in calibration, while the frequency-only model performs substantially worse than the full architecture. The time-only model remains comparatively competitive in this particular test, but the full time-frequency model with contrastive learning consistently provides the most reliable calibration across the architectures considered. These results show that the proposed framework improves not only the central location of the recovered posterior, but also the statistical reliability of its credible intervals under strong transient contamination.

Taken together, Figs.~\ref{fig:triangle} and \ref{fig:ppplot} demonstrate that the proposed method achieves two properties simultaneously in contaminated data: accurate posterior localization and robust calibration. This is an important distinction, because a posterior may exhibit a narrow credible region while remaining statistically miscalibrated. Here, both diagnostics favor the proposed framework over the conventional MCMC baseline, indicating that the learned model successfully mitigates the bias introduced by transient glitches while preserving meaningful posterior uncertainty.

A further practical advantage of the proposed framework is its amortized nature. Once training has been completed, posterior inference for a new event requires only a forward pass through the encoder and conditional flow, avoiding the high per-event computational cost of iterative MCMC sampling. For the representative glitch-contaminated case considered here, generating $3.2\times10^4$ posterior samples with the conventional MCMC baseline requires 23~min~25~s, whereas the trained NPE model produces the same number of samples in $0.6~\mathrm{s}$, with both measurements obtained on an NVIDIA RTX A6000 GPU with 48~GB memory. This makes the method especially attractive for rapid analysis in future space-based GW observations, where low-latency source characterization may be important for coordinated astrophysical follow-up.

\subsection{Architectural contribution of multimodal fusion and contrastive learning}
\label{subsec:results_ablation}

We next investigate the architectural factors responsible for the robustness gains observed in the full model. To this end, we perform systematic ablation studies comparing the complete time-frequency model with contrastive learning against reduced variants that remove either the multimodal design or the contrastive objective.

\begin{figure}[t]
    \centering
    \includegraphics[width=0.98\columnwidth]{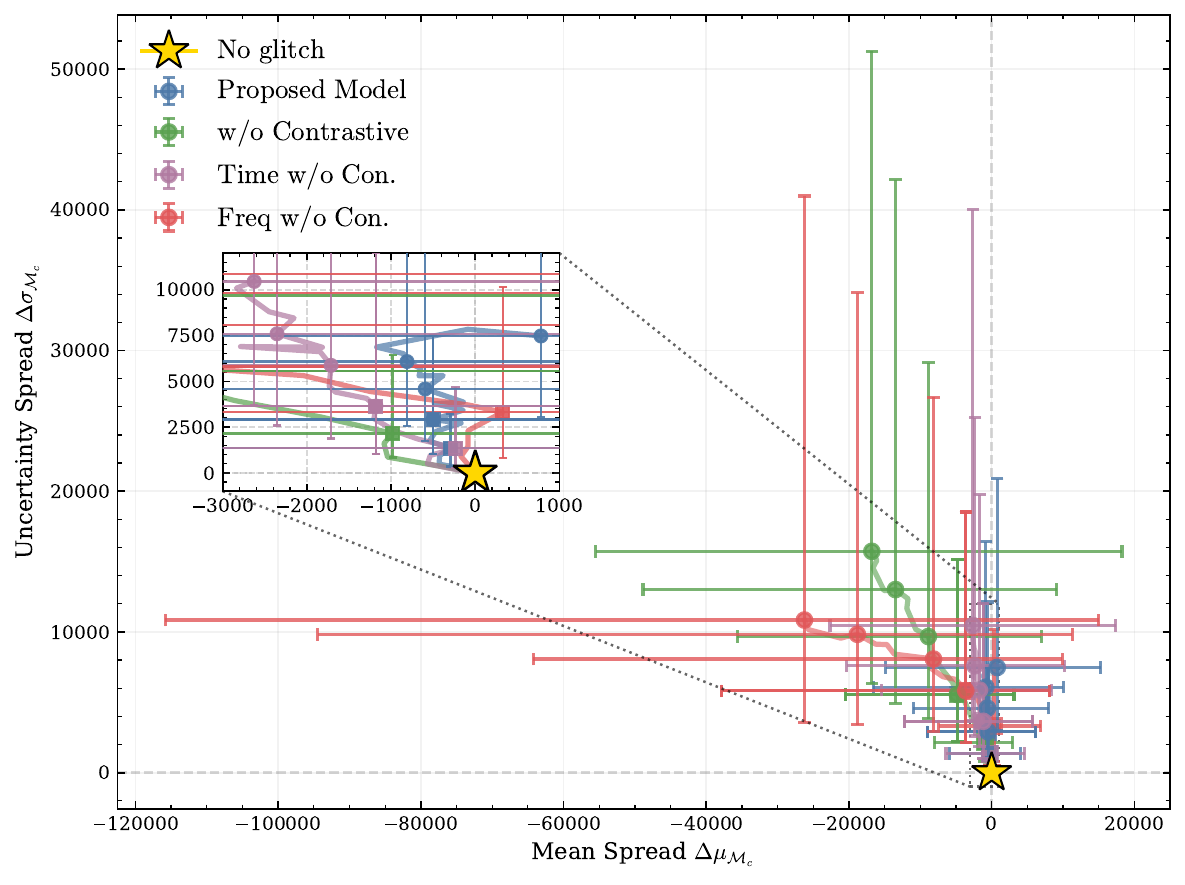}
    \caption{Spread-statistics trajectory analysis of posterior accuracy and uncertainty across model variants. The horizontal axis represents the mean posterior offset $\Delta~\mu_{\mathcal{M}_c}$ relative to the uncontaminated reference, and the vertical axis represents the corresponding uncertainty spread $\Delta~\sigma_{\mathcal{M}_c}$. Each point denotes the behavior of one model variant under contaminated conditions, and the SNR of the glitch gradually increases along the direction of the line.}
    \label{fig:trajectory}
\end{figure}

Figure~\ref{fig:trajectory} shows the spread-statistics trajectories for different model variants relative to the uncontaminated reference. The full model remains most tightly clustered around the low-deviation region, indicating the most favorable combination of small posterior mean offsets and controlled uncertainty widths. In contrast, removing the contrastive objective or restricting the model to a single representation domain leads to visibly larger excursions away from the clean-reference regime. These trajectories show that the full architecture does not merely improve one aspect of inference at the expense of another, but achieves a more balanced performance in terms of both accuracy and spread.

\begin{figure}[t]
    \centering
    \includegraphics[width=0.48\textwidth]{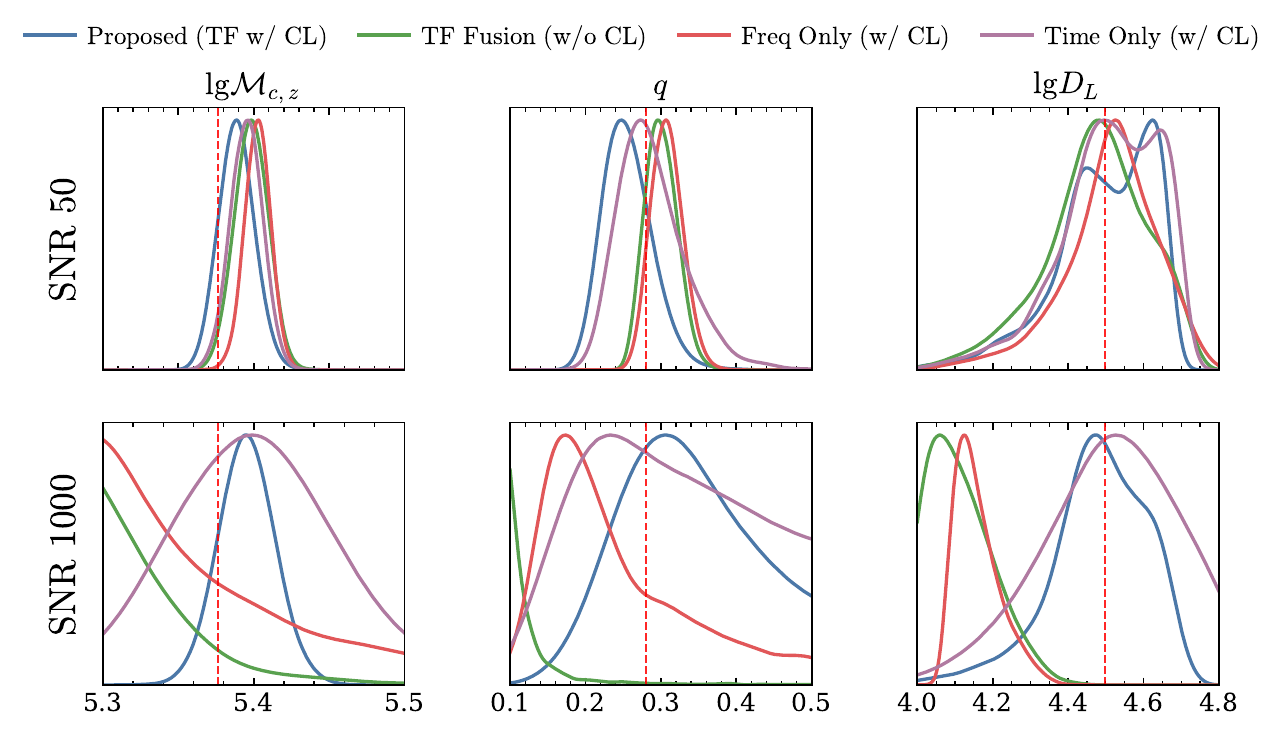}
    \caption{Comparison of marginalized posteriors across model architectures and contamination levels. The panels show one-dimensional marginalized posterior distributions for representative source parameters, including $log{\mathcal{M}_c,z}$, $q$, and $log{D_L}$, under different glitch contamination regimes. Curves correspond to the proposed full model, the time-frequency model without contrastive learning, and the time-only and frequency-only variants.}
    \label{fig:snrcompare}
\end{figure}

This trend is further corroborated by the marginal posterior comparisons in Fig.~\ref{fig:snrcompare}, where the full proposed method is evaluated against alternative baseline architectures across multiple contamination levels. Under more moderate conditions, the reduced models may still recover broadly reasonable posterior shapes, but their marginals already begin to show larger distortions, broader spreads, or shifts relative to the target structure. As the contamination becomes more severe, these differences become increasingly evident: the full time-frequency model with contrastive learning remains closest to the desired posterior behavior, whereas the single-domain and non-contrastive variants degrade more visibly.

The ablation results support two main conclusions. First, time-domain and frequency-domain representations provide genuinely complementary information for glitch-robust inference. The time domain is particularly informative for transient morphology, amplitude envelopes, and localized disturbances, whereas the frequency domain captures chirping signal structure and the spectral imprint of contamination. A model restricted to only one of these domains necessarily discards useful information that becomes especially important under non-stationary conditions. Second, the contrastive objective is not merely a marginal regularization term, but a primary contributor to inference robustness. By encouraging different contaminated realizations of the same source to map to nearby latent representations, it actively suppresses nuisance variability and stabilizes the posterior estimator against glitch-induced distortions.

Thus, Figs.~\ref{fig:trajectory} and \ref{fig:snrcompare} together show that the strongest performance does not arise from multimodal fusion alone or from contrastive learning alone, but from their combination. The full model benefits both from access to complementary time-frequency information and from an explicit representation-level mechanism that encourages invariance to transient nuisance structure. This combined design is what enables the robust behavior observed in the earlier inference and calibration tests.

\subsection{Robustness to glitch duration and merger-relative timing}
\label{subsec:results_robustness}

\begin{figure}[t]
    \centering
    \includegraphics[width=0.48\textwidth]{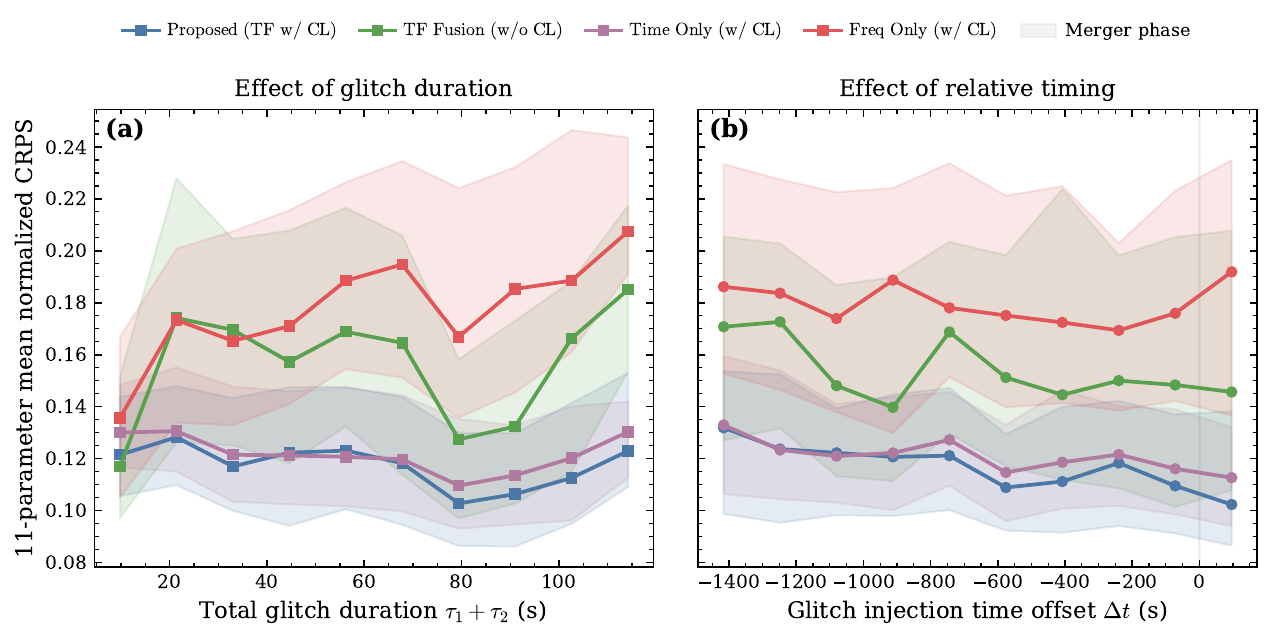}
    \caption{Robustness of inference performance to glitch duration and merger-relative timing. Panel (a) shows the mean normalized CRPS over the full 11-parameter space as a function of the total glitch duration $\tau_1 + \tau_2$. Panel (b) shows the same metric as a function of the glitch injection time offset relative to the binary merger. Different curves correspond to the proposed full model and the reduced baseline architectures. Lower CRPS indicates better posterior fidelity.}
    \label{fig:durationoffset}
\end{figure}

Finally, we investigate whether the performance of the trained models depends strongly on the temporal properties of the contaminating glitch. Figure~\ref{fig:durationoffset} summarizes the mean normalized CRPS over the full 11-parameter space as a function of two quantities: the total glitch duration and the time offset of the glitch relative to the binary merger. These scans provide a more stringent robustness test because they probe whether the model remains reliable when the contamination geometry changes in a systematic way.

For the proposed framework, the CRPS remains comparatively stable over the explored range of both variables. This indicates that inference quality does not deteriorate sharply as the glitch becomes longer or as its location shifts relative to the merger time. Within the tested regime, neither quantity emerges as a dominant failure mode for the full model. In other words, the learned representation appears capable of handling a broad family of overlaps between the astrophysical signal and the transient artifact.

This behavior stands in contrast to the stronger morphology dependence often encountered in conventional MCMC analyses, where inference quality can vary significantly depending on the precise temporal overlap between the signal and the non-stationary disturbance. In the present framework, robustness is learned directly from large ensembles of contaminated realizations rather than imposed through a rigid analytical likelihood. As a result, the network can adapt to many contamination patterns without requiring case-specific analytical treatment.

The comparison across architectures in Fig.~\ref{fig:durationoffset} also shows that this stability is not a generic property of any neural model trained on contaminated data. The full time-frequency model with contrastive learning consistently achieves the lowest or near-lowest CRPS across the scans, whereas the reduced variants show higher scores and stronger sensitivity to the contamination geometry. This further supports the conclusion that the combination of multimodal fusion and contrastive learning is central to the observed robustness.

\section{Conclusion}
\label{sec:conclusion}

In this work, we developed a glitch-robust parameter inference framework for massive black hole binaries in Taiji space-based  GW observations. The method combines simulation-based amortized inference with a conditional normalizing flow, a time-frequency multimodal fusion encoder, and a contrastive learning objective designed to suppress transient nuisance information. To support large-scale training, we further introduced a neural glitch generator that produces high-fidelity synthetic transients at substantially reduced computational cost.

Systematic experiments demonstrate that the proposed framework achieves reliable posterior recovery even in the presence of strong non-stationary glitch contamination. Relative to a conventional MCMC baseline, the method yields posteriors that remain better aligned with the injected source parameters and retain significantly improved calibration under contaminated conditions. Among the neural architectures examined, the full time-frequency model with contrastive learning consistently delivers the strongest overall performance, showing that both multimodal fusion and contrastive learning are essential for robust inference rather than optional refinements.

We also find that the trained framework remains stable, within the explored range, against variations in glitch duration and merger-relative timing. This suggests that the dominant limitation of traditional approaches arises less from an intrinsic degeneracy of the data themselves than from the mismatch between glitch-contaminated detector data and stationary analytical likelihood assumptions. By learning directly from contaminated simulations, the proposed framework avoids this mismatch and remains effective across a broad variety of glitch morphologies.

Beyond the specific architecture considered here, the present study also highlights an important methodological point for posterior validation. Standard marginal coverage diagnostics such as P-P plots are necessary for assessing calibration, but they do not fully characterize the agreement between the recovered posterior shape and the target distribution. For this reason, we complement coverage-based validation with the CRPS, which provides a more informative measure of global distributional fidelity. This combined evaluation strategy should be broadly useful for Bayesian inference in non-ideal  GW data.

Future work should extend the framework to more diverse and weakly modeled glitch populations, including overlapping or multi-glitch scenarios, and explore transferability to broader classes of space-based sources. In particular, the present study assumes a relatively controlled simulation setting in which glitches are injected on top of stationary Gaussian noise, whereas real instrumental disturbances may be accompanied by concurrent changes in the noise level, spectral structure, or broader non-stationary detector environment. Addressing these effects would further strengthen the case for deep-learning-based amortized inference as a practical analysis paradigm for next-generation  GW observatories. Taken together, these results suggest that deep-learning-based amortized inference can serve not merely as a computational surrogate for traditional samplers, but as a practically viable Bayesian analysis paradigm for space-based  GW observations under non-ideal data conditions.

\section*{Acknowledgements}
This work was supported by the National Natural Science Foundation of China (Grants Nos. 12473001, 12575049, and 12533001), the National SKA Program of China (Grants Nos. 2022SKA0110200 and 2022SKA0110203), the China Manned Space Program (Grant No. CMS-CSST-2025-A02), and the 111 Project (Grant No. B16009).

\section*{Data Availability}
The data supporting the results of this article will be made publicly available in an online repository upon publication.

\bibliography{glitch_taiji}

\appendix
\section{Methodological Scope and Degeneracies of Coverage Testing}
\label{app:coverage_limitations}

This appendix clarifies an important limitation of standard marginal coverage tests, such as P-P plots evaluated with the KS statistic. Although widely used to assess calibration, passing a marginal coverage test does not inherently guarantee that the recovered posterior distribution captures the correct overall morphology. This limitation is not specific to the present model, but is a general feature of Bayesian posterior validation based only on one-dimensional coverage.

To illustrate this point, we consider a simplified statistical experiment that isolates two basic forms of posterior distortion: a systematic mean shift $\Delta\mu$ and a variance scaling factor $\alpha$. Instead of repeatedly rerunning full end-to-end waveform inference, we draw true parameter values from a standard normal distribution and compare them with predicted posterior distributions whose mean and variance are artificially modified. This construction provides a computationally efficient way to map how the validation metrics respond to controlled distortions in posterior location and width. Because both the KS statistic and the CRPS depend on the relative position of the true value within the predicted posterior, this simplified setup captures the essential behavior relevant for posterior validation.

\begin{figure}[htbp]
    \centering
    \includegraphics[width=0.48\textwidth]{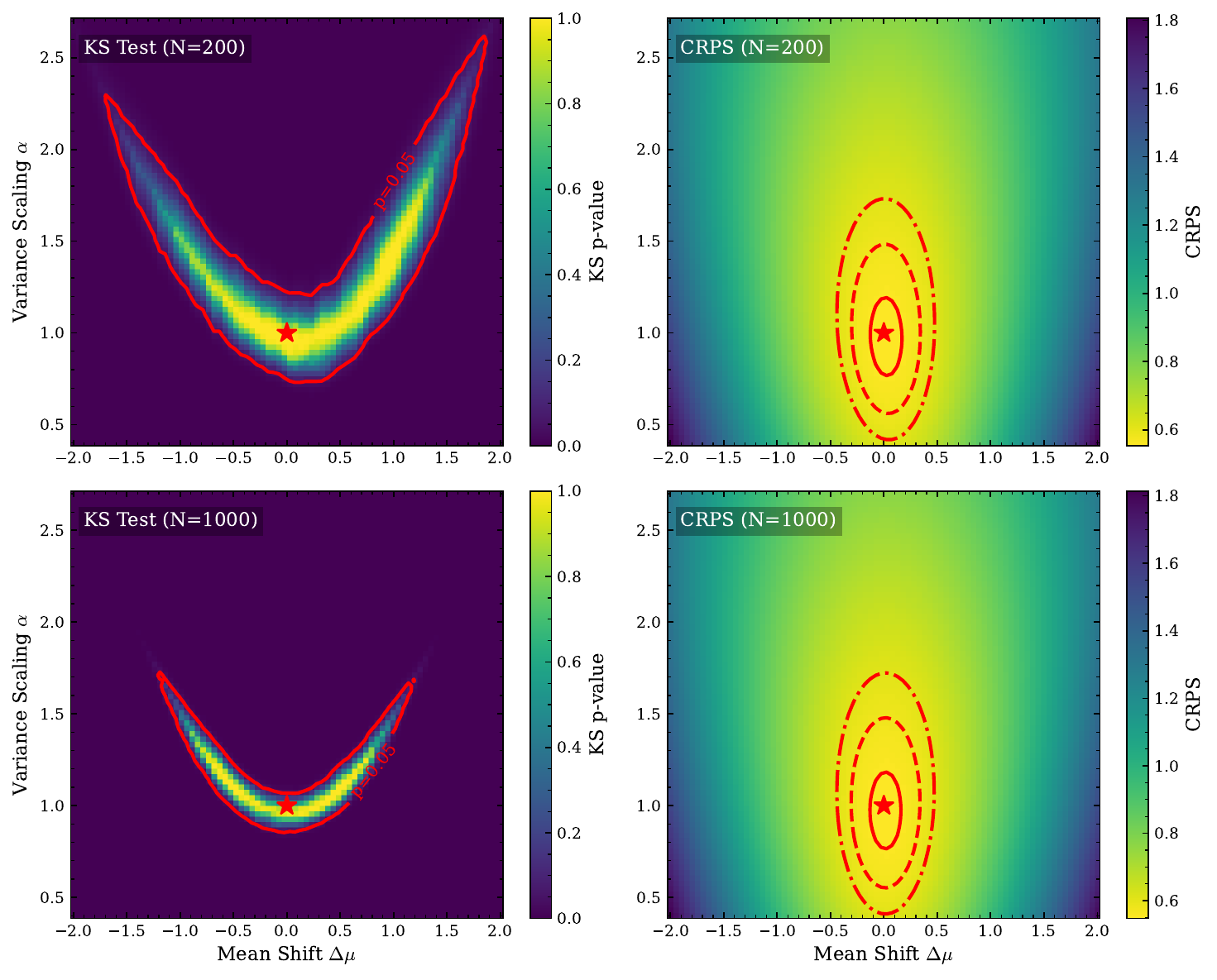}
    \caption{Illustration of marginal-coverage diagnostics in a simplified relative-shift experiment. The left panel shows the KS-test p-values in the $(\Delta\mu,\alpha)$ plane, and the right panel shows the corresponding CRPS values. The nominally consistent configuration is located near $(\Delta\mu=0,\alpha=1)$.}
    \label{fig:kscrps}
\end{figure}

The resulting metric landscape is shown in Fig.~\ref{fig:kscrps}. The KS test exhibits an extended region in the $(\Delta\mu,\alpha)$ plane for which distorted posterior distributions still yield acceptable p-values. In other words, a family of shifted or rescaled posteriors can remain statistically consistent with the expected marginal coverage, even though their shapes differ from the target distribution. Passing the KS-based coverage test therefore confirms that the nominal credible intervals contain the true values at approximately the expected rates, but it does not ensure that the posterior morphology is fully correct, especially in a broader distributional sense~\cite{DBLP:journals/tmlr/FerrerR25,DBLP:journals/corr/abs-2504-11299}.

To complement this limitation, we additionally consider the CRPS, which is a strictly proper scoring rule sensitive to both location and scale errors. Unlike the KS landscape, which admits a broad degenerate valley of passing posterior distortions, the CRPS landscape penalizes both mean offsets and variance miscalibration more directly. As shown in Fig.~\ref{fig:kscrps}, the CRPS minimum is localized near the theoretically consistent point $(\Delta\mu=0,\alpha=1)$, thereby providing a stronger constraint on posterior shape fidelity.

For this reason, the main text adopts CRPS as a complementary metric alongside standard coverage diagnostics. In our framework, P-P plots and KS tests remain useful for assessing marginal calibration, but CRPS is required to probe whether the recovered posterior also matches the target distribution in a stronger distributional sense. The combination of these two diagnostics provides a more reliable evaluation protocol for amortized posterior estimators in non-ideal  GW data analysis.

\end{document}